\documentclass[letter,twocolumn]{jpsj3} 
\usepackage{graphicx}
\usepackage{dcolumn}
\usepackage{bm}
\usepackage{amsmath}
\usepackage{amsfonts}
\usepackage{amssymb}%
\usepackage{color}%

\providecommand{\U}[1]{\protect\rule{.1in}{.1in}}
\title{
{{\it Ab initio} Studies {of} 
Magnetism in {the} Iron Chalcogenides FeTe and FeSe}
}

\author{
Motoaki Hirayama\thanks{E-mail: hirayama@stat.phys.titech.ac.jp},$^{1}$ Takahiro Misawa,$^{2}$ Takashi Miyake,$^{3}$ and Masatoshi Imada$^{2}$
}
\inst{
$^1$Department of Physics, Tokyo Institute of Technology, Ookayama, Meguro-ku, Tokyo 152-8551, Japan 
$^2$Department of Applied Physics, University of Tokyo, 7-3-1 Hongo, Bunkyo-ku, Tokyo 113-8656, Japan
$^3$Nanomaterials Research Institute, AIST, Tsukuba{, Ibaraki} 305-8568, Japan
}

\date{\today}

\abst{
{The iron chalcogenides} FeTe and FeSe belong to the family of iron-based superconductors.
We study {the} magnetism {in} these compounds in the normal state using the {\it ab initio} downfolding
scheme developed for strongly correlated electron systems.
In deriving $ab$ $initio$ low-energy effective models,
we employ the constrained $GW$ method to eliminate the double counting of electron correlations
originating from the exchange correlations already taken into account in the density functional theory.
By solving the derived {\it ab initio} effective models, we reveal that
the elimination of the double counting is important in reproducing 
the bicollinear antiferromagnetic order in FeTe, as is observed in experiments.
We also show that the elimination of the double counting induces 
a {unique} degeneracy of several magnetic orders in FeSe, which may explain the absence of the magnetic ordering.
We discuss {the relationship} between the degeneracy and the recently found puzzling phenomena
in FeSe {as well as the magnetic ordering found under pressure}.
}

\begin{document}
\maketitle


After the discovery of iron-based superconductors, many families of
the iron-based superconductors have been found~\cite{IshidaJPSJReview,StewartRMP2011,ScalapinoRMP}. 
Among them, iron chalcogenides such as FeTe and FeSe (11-type) 
are the simplest iron-based superconductors, which 
have no blocking insulating layer.
{Despite} the simple lattice structure of the 11-type compounds,
many characteristic behaviors have been 
observed in {them}, {which {have}} attracted much interest.
For example, 
FeTe shows an exceptional bicollinear antiferromagnetic (AFB) order
with {an} ordered moment {of} $\sim 2.0$--$2.25 \mu_{\text{B}}$~\cite{FeTe_PRB2009, bao09},
although most of the iron-based superconductors 
show {a} stripe antiferromagnetic (AFS) order.
Although the AFB order itself is
reproduced by
several calculations based 
on the density functional {theory} (DFT)~\cite{ma09,moon10,kumar12,YZZhangPRB2013},
{the} microscopic origin of the AFB order and 
{the} roles of electron correlations
are not fully clarified yet. 

In FeSe, although calculation based on the DFT
predicts that the AFS order is the ground state~\cite{kumar12},
no clear signature of magnetic order
is observed in {experiments,} and the superconducting 
phase appears below $T_{c}\sim${8 K}~\cite{FeSeHsu2008}.
{Note} that a structural phase transition without
magnetic {order} occurs at {$T_{s}\sim${90 K}}. 
The absence of the magnetic order {despite} its large local moment~\cite{FeSeLocalMoment} {remains} a puzzle in FeSe.
Recent {experimental result}~\cite{TerashimaPRB2014,ShimojimaPRB2014,Kasahara2014} 
supports {the notion} that FeSe has extremely small Fermi surfaces and 
{that the} resultant superconductivity is located in the
Bose-Einstein-condensate (BEC) region.
{The} origin of these exotic phenomena {also remains} a puzzle in FeSe.
Furthermore, it is reported that {a} single layer of FeSe on SrTiO$_{3}$
shows high-temperature superconductivity, 
whose $T_{c}$ is higher than {50 K}~\cite{Qing_2012}.
Thus, clarifying these {remarkable} behaviors 
in iron chalcogenides is important {for understanding} the nature 
of high-temperature superconductivity in iron-based superconductors.

Many theoretical and experimental studies indicate that 
{the} 11-type compounds are located in the strong{-}coupling region~\cite{nakayama10,miyake2010,Aichhorn,ZPYin,Lanata_PRB,FeSeLocalMoment,TerashimaPRB2014,ShimojimaPRB2014,Kasahara2014}. 
{In particular}, in recent experimental studies {of} a single crystal of FeSe~\cite{TerashimaPRB2014,ShimojimaPRB2014,Kasahara2014},
it {has been} shown that the experimental Fermi surfaces {cannot} be explained by
the conventional DFT calculations.
In addition, an $ab$ $initio$ evaluation of the interaction parameters 
suggests that the 11-type compounds have relatively stronger correlations
{than} other iron-based superconductors such 
as BaFe$_{2}$As$_{2}$ and LaFeAsO~\cite{miyake2010}.
In {the} strong{-}coupling region, 
it is necessary to consider the 
effects of {electron correlations} beyond the 
conventional DFT.
Even when the strong {electron correlation} is taken into account,
the above {recently studied} Fermi surface does not appear consistent with the calculated Fermi surface~\cite{Aichhorn,JanPRL2012}.
This result is a puzzling aspect of FeSe.

A hybrid method ($ab$ $initio$ downfolding scheme)~\cite{ImadaMiyake} of the DFT and 
precise model calculation 
is a promising way of treating electron correlations
beyond the standard DFT:
In this scheme,
we first obtain the global band structure based on the DFT and 
evaluate the interaction parameters 
in the low-energy effective model by
constrained random-phase approximation (cRPA)\cite{aryasetiawan04,miyakeAryasetiawan08}.
Then, we solve the low-energy effective model using 
precise low-energy solvers such as {the} many-variable variational Monte Carlo (mVMC) method~\cite{TaharaVMC_Full}.
Several applications to real materials 
show the validity and accuracy of this scheme~{\cite{misawa2011,misawa2012,shinaoka12}}.
However, in this scheme,
there is a fundamental 
problem {in} that the {electron correlations} are doubly counted, i.e.,
the {electron correlations} are considered 
 in both the band calculations and the model calculations.
Although this double counting is assumed to be small in 
the conventional downfolding scheme,
in the strong-coupling regime where 
iron chalcogenides are expected to be located, 
{the} elimination of double counting 
{may become} important to correctly reproduce electronic structures.

In this Letter, 
by employing the state-of-the-art $ab$ $initio$ method,
we clarify the magnetic properties in iron chalcogenides.
We properly eliminate the  double counting of 
the {electron correlations} and derive the
low-energy effective models in an $ab$ $initio$ way~\cite{hirayama13}.
Then, we solve the low-energy effective models using the mVMC.
As a result, we show that the elimination of {the} double counting plays a crucial role in
stabilizing the AFB in FeTe.
{W}e also solve the low-energy effective models
for FeSe {in addition to FeTe}, and find that several magnetic orders are energetically 
nearly degenerate.
This degeneracy may result in the 
absence of the magnetic order in FeSe under competitions and frustrations emerging in the region close to the phase boundaries.

\begin{figure}[tb]
\centering  
\includegraphics[width=0.45\textwidth ]{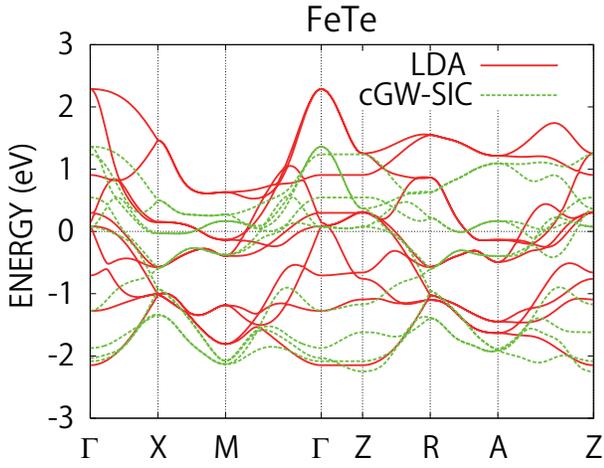}  
\caption{
(Color online) Electronic band structures of the Fe $3d$ {MLWFs} of FeTe in the LDA [(red) solid line]
 and the {cGW-SIC} [(green) dashed line].
The Fermi energy is set to zero.
}
\label{FeTe_band}
\end{figure} 

Our low-energy effective models for iron-based superconductors are given as follows:
\begin{multline}
\mathcal{H}_{\text{eff}} ^{\text{X}}= \sum_{ij} \sum_{mn\sigma }
t^{\text{X}}_{mn\sigma}(\bm{R}_i-\bm{R}_j) d_{in\sigma} ^{\dagger} d_{jm\sigma} + \frac{1}{2} \sum_{ij} \sum_{mn \sigma \rho} \\ 
\biggl\{ U_{mn\sigma \rho }(\bm{R}_i-\bm{R}_j) d_{in\sigma}^{\dagger}  
d_{jm\rho}^{\dagger} d_{jm\rho} d_{in\sigma} +J_{mn\sigma \rho}(\bm{R}_i-\bm{R}_j) \\
 \bigl( d_{in\sigma}^{\dagger} d_{jm\rho}^{\dagger} d_{in\rho} d_{jm\sigma} 
+d_{in\sigma}^{\dagger} d_{in\rho}^{\dagger} d_{jm\rho} d_{jm\sigma}\bigr) \biggr\},
\label{Hamiltonian}  
\end{multline}
\begin{equation}
t^{\text{X}}_{mn\sigma}(\bm{R})= \langle \phi _{m\bm{0}}|\mathcal{H}^{\text{X}}|\phi _{n\bm{R}} \rangle, 
\label{t}
\end{equation}
where $t^{\text{X}}$  
represents {the} transfer integral of the maximally localized Wannier functions ({MLWFs})~\cite{marzari97,souza01}.
Here, {$\text{X}$} denotes the methods  
{of} deriving {the} one-body part of the Hamiltonian, i.e.,
 {$\text{X}$}=LDA (cGW-SIC) means that the local density approximation (constrained GW-self-interaction correction)
is used for evaluating transfer integrals.
We will show details of {the} cGW-SIC later.
{The} corresponding band structures near the Fermi level are shown
in Fig.~\ref{FeTe_band}.
Here, $d_{in\sigma} ^{\dagger}$ ($d_{in\sigma}$) 
is a creation (annihilation) operator of an 
electron with spin $\sigma$ in the $n$th MLWF having strong Fe $3d$ characters (with $n=1$--$5$ being the orbital index) centered at $\bm{R}_{i}$~\cite{miyake2010}.
In this {work}, we take {the} X- and Y-{axes} for the {MLWFs} index along the Fe-Te/Se directions.
{Note} that the $12\times 12\times 6$ $\bm{k}$-mesh is employed in the LDA calculation, 
and the $6\times 6\times 3$ $\bm{k}$-mesh is employed in the cRPA and cGW-SIC calculations.
{The} interaction parameters $U$ and $J$ denote the Coulomb 
interactions and exchange interactions, respectively.
These interaction parameters 
are calculated by the cRPA~\cite{aryasetiawan04,ImadaMiyake}
and dimensional downfolding~\cite{nakamura2010,nakamurap,misawa2012}.
For details of {the} interaction parameters and hopping parameters, see {Ref.~\citen{SM}}.
The off-site interactions ($i\neq j$) are omitted
in this calculation because they are less than {one-}quarter of the on-site interaction parameter. 
Double counting in the Hartree term is subtracted 
as we have mentioned in our previous paper~\cite{misawa2011}.

For the one-body part of the Hamiltonian, in addition to the standard LDA calculations,
{the} constrained GW (cGW) scheme {was employed}~\cite{hirayama13,aryasetiawan09}, 
which can eliminate the double counting of the 
self-energy originating 
from the exchange-correlation energy.
In this scheme,
we first subtract the exchange-correlation energy 
in the LDA calculations and replace it with the $GW$ self-energy
{calculated by excluding the low-energy contribution},
since {the low-energy part of the self-energy} will be considered later when the obtained model is solved.
In this scheme, since we do not consider the low-energy part of the 
$GW$ self-energy in deriving the low-energy effective model, 
the double counting of {exchange correlations} does not occur.

The cGW self-energy does not include the self-interaction (SI) originating from the low-energy degree of freedom,
which should be canceled out {by} the opposite-sign SI in the Hartree term.
In the {nondegenerate multiband} systems, the imbalance of the 
SI {exerts} crucial effects on the occupation number of each orbital and the magnetism.
We therefore introduce the self-interaction correction (SIC) originating from 
the low-energy degrees of freedom $U^{\text{on-site}} n^{L}_{\text{LDA}}/2 $ {into} the cGW scheme, 
where $U^{\text{on-site}}$ is the on-site effective interaction 
and $n^{L}_{\text{LDA}}/2$ is the occupation number of the 
low-energy degrees of freedom of the up or down spin in the LDA.
The one-body part with the SIC is given by
\begin{equation}
\mathcal{H}^{\text{cGW-SIC}}=\mathcal{H}^{\text{cGW}}-Z^{\text{cGW}}U^{\text{on-site}} \frac{n^{L}_{\text{LDA}}}{2}  ,
\label{HcGW-SIC}
\end{equation}
where $\mathcal{H}^{\text{cGW}}$ is the static one-body Hamiltonian in the cGW~\cite{hirayama13}.
The renormalization factor $Z^{\text{cGW}}$ is needed to renormalize 
{the} frequency{-}dependent part of the interaction of the effective model~\cite{hirayama13}.

Figure \ref{FeTe_band}  shows the band structures 
of FeTe in the LDA and cGW-SIC,
which correspond to the one-body part of 
the low-energy effective model in Eq. (\ref{Hamiltonian}).
We note that the band {structure} of the cGW-SIC model for {the} noninteracting case 
is a virtual one, which cannot be directly observed in experiments.
It is expected that correct band structures are reproduced 
after solving the low-energy effective models.
The main difference between the LDA and cGW-SIC models is the
on-site potential of the $3Z^2-R^2$ orbital.
In the cGW-SIC model,
the on-site potential of the $3Z^2-R^2$ orbital decreases
and the occupation of the $3Z^2-R^2$ orbital increases.
As we {will} show later, {the} stability of the magnetic order largely depends on
the on-site potential of the $3Z^2-R^2$ orbital.

To solve the low-energy effective models, we employ the mVMC, which 
can treat the {electron correlations} seriously~\cite{TaharaVMC_Full}.
Our variational wavefunction is defined as
\begin{equation}
|\psi \rangle = \mathcal{P}_{\text{J}}\mathcal{P}_{\text{G}}\mathcal{L}^{S=0}|\psi _{\text{pair}} \rangle ,
\end{equation}
where $|\psi _{\text{pair}} \rangle $ is a 
generalized pairing wave function, 
and $\mathcal{L}^{S=0}$, $\mathcal{P}_{\text{G}}$, 
and $\mathcal{P}_{\text{J}}$ are the total spin quantum-number projection, 
Gutzwiller factors,
and Jastrow factors,
respectively~\cite{TaharaVMC_Full,misawa2011}.
For more computational details {of} iron-based superconductors,
readers are referred to {Ref.~\citen{misawa2011}}.

\begin{figure}[t]
\centering 
\includegraphics[clip,width=0.4\textwidth ]{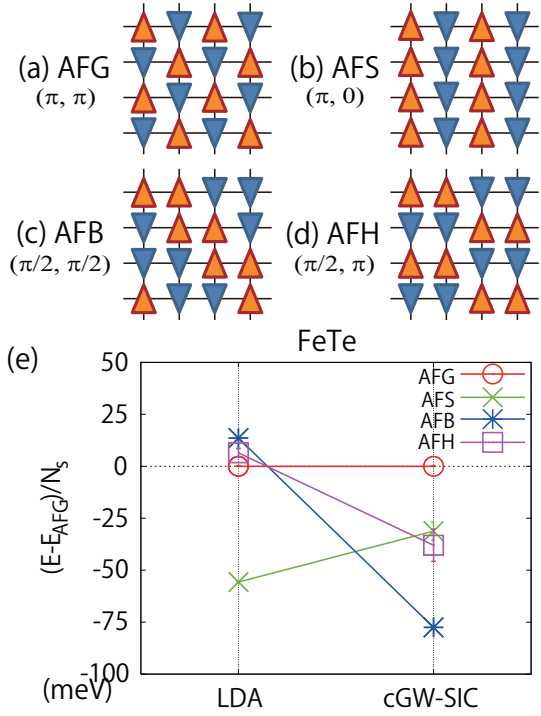} 
\caption{(Color online) AF patterns calculated in this {work}: (a) AFG, (b) AFB, (c) AFS, and (d) AFH. 
(e) {Groundstate} energy per site of AF in mVMC extrapolated to the thermodynamic limit 
for FeTe with $t^{\text{LDA}}$ and with $t^{\text{cGW-SIC}}$ (in meV). 
The energy $E$ is extrapolated to the thermodynamic limit $E_{\infty}$ by employing the 
scaling $E=E_{\infty} + aN_{\text{s}}^{-3/2}$~\cite{HusePRB88,shinaoka12} with a constant $a$ 
for the system size $N_{\text{s}}$.
}
\label{AF_FeTe_energy}
\end{figure}

In the mVMC, we examine the stability of four possible AF orders, i.e.,
G-type AF state (AFG), AFS, AFB, and half-collinear 
AF state (AFH), whose schematic pictures are
{shown} in {Figs.~\ref{AF_FeTe_energy}(a)-\ref{AF_FeTe_energy}(d), respectively}.
We mainly show the results of $N_{\text{s}}=8\times 8$ sites
with periodic boundary {conditions;}
the systems with $4\times 4$ and $4\times 8$ 
sites are also calculated for the extrapolation to the thermodynamic limit.

By solving the low-energy effective models, 
we show how the elimination of double counting affects the 
stability of magnetic orders in FeTe. 
As shown in {Figs.~\ref{AF_FeTe_energy}(a)-\ref{AF_FeTe_energy}(d)},
we examine four different magnetic orders, which 
are potential candidates of the order.
In the LDA model,
we find that the AFS is the most stable {magnetically} ordered phase {[Fig.~\ref{AF_FeTe_energy}(e)]}.
In contrast, the AFB phase becomes 
the ground state in the cGW-SIC {calculation} that properly eliminates the 
double counting of the self-energy. This result indicates the 
importance of the elimination of {the} double counting 
in the strong-coupling regime.

In Table~\ref{Occ_VMC}, we show the orbital resolved occupation number $n_{\nu}$
for both the LDA model and the cGW-SIC model.
Compared {with that in} the LDA model, the occupation 
number of the $YZ/ZX$ orbitals becomes close to
half filling ($n_{YZ/ZX}\sim 1$) in addition to the 
$X^2-Y^2$ orbital.
This is caused by the lowering of {the} on-site potential of the $3Z^2-R^2$ orbital
in the cGW-SIC model.
The localization of the $X^2-Y^2$ orbital is frequently 
referred {to} as the orbital selective Mottness 
in the literature~\cite{IshidaLiebsch,Aichhorn,misawa2011,misawa2012,Greger_2013,MediciCapone}.
In addition, our result indicates that the 
$YZ/ZX$ orbitals also show orbital{-}selective Mott insulating behavior 
in FeTe as well as {in} FeSe.
The orbital selective Mott insulating behaviors are evidenced by small double occupancies {$d_m=\langle n_{m\uparrow}n_{m\downarrow} \rangle =0.1$ for $m=$} $YZ/ZX$ {and $=0.04$ for $m=$} $X^2-Y^2$
and {small} orbital-{resolved} charge compressibilities
{$\chi_{\text{c}m}= d n_{m}/d \mu  <0.001$  (meV)$^{-1}$  for $m=$
$YZ/ZX$ and $X^2-Y^2$  in FeTe (AFB)
({$\mu = d E/d N$} is {the uniform} chemical potential {applied to all the orbitals}{, where $E$ is the total energy and $N$ is the total number of electrons).}}~\cite{MisawaNC14}.
Because {the} occupancies of the $YZ/ZX$ orbitals are slightly different from half filling,
the $YZ/ZX$ orbitals can be regarded as doped orbital{-}selective Mott insulators in a strict sense,
where the excess carriers from half filling make the system metallic~\cite{MizuguchiJPSJ10}.
This additional localization may be the origin of the characteristic 
behavior observed in iron chalcogenides.
{Note} that {a} similar localization  is observed in previous 
calculations for iron chalcogenides~\cite{Lanata_PRB,MediciCapone}.

\begin{table}[tb] 
\caption{Occupation {numbers} of the 
Wannier functions of the {LDA and cGW-SIC models} in the mVMC, 
where the sum of the occupancy is $6$.
Error bars are {negligible} ($\leq  10^{-3}$).
}
\
\begin{tabular}{c|ccccc}
\hline \hline \\ [-8pt]  
FeSe (AFH)               & $XY$ & $YZ$ & $3Z^2-R^2$ & $ZX$ & $X^2-Y^2$   \\ [+1pt]
\hline \\ [-8pt] 
LDA                      & 1.14 & 1.36 & 1.02 & 1.32 & 1.16    \\
cGW-SIC                  & 1.27 & 1.04 & 1.64 & 1.04 & 1.01    \\
\hline \\ [-8pt]
FeTe (AFB)               & $XY$ & $YZ$ & $3Z^2-R^2$ & $ZX$ & $X^2-Y^2$   \\ [+1pt]
\hline \\ [-8pt] 
LDA                      & 1.40 & 1.12 & 1.12 & 1.31 & 1.05    \\ 
cGW-SIC                  & 1.44 & 1.05 & 1.42 & 1.07 & 1.02    \\  
\hline \hline 
\end{tabular}
\label{Occ_VMC} 
\end{table}

\begin{figure}[tb]
\centering 
\includegraphics[clip,width=0.4\textwidth ]{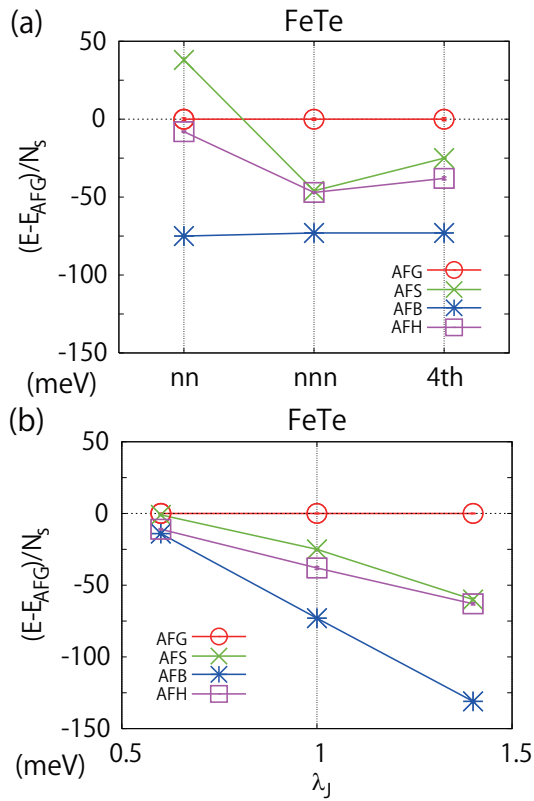} 
\caption{
(Color online) Ground-state energy per site of AF in the mVMC for FeTe with the cGW-SIC parameter
(a) as a function of cutoff length of $t$
and (b) as a function of ratio for $J$, $\lambda _{J}$ ($J \rightarrow \lambda _{J}\cdot J$) (in meV).
{The} nearest-neighbor, next-nearest-neighbor, and $4$th-neighbor hoppings are included in "nn", "nnn", and "4th" calculations, respectively.
Energy is measured from that of the AFG.
{The} system size is $N_{\text{s}}=8\times 8$.
}
\label{FeTe_energy_Jtcut}
\end{figure}

Here, we examine the microscopic origin of the 
bicollinear magnetic order in FeTe.
In previous calculations based on {the} DFT~{\cite{ma09,moon10}},
although it was suggested that the origin of the AFB order is attributed to
large farther-neighbor magnetic interactions such as 
the third-neighbor magnetic interaction $J_3$,
it is unclear what induces the large {farther}-neighbor magnetic interactions.
One possible origin of
such large farther-neighbor interactions is the 
superexchange interaction that is
induced by farther-neighbor hoppings.~To examine 
the effects of farther-neighbor hoppings,
we artificially cut off farther-neighbor hoppings. 
As shown in Fig.~\ref{FeTe_energy_Jtcut}(a),
we find that the AFB becomes {the} ground state {regardless} of the range of 
such hoppings.
The robustness of the AFB indicates {that} the 
superexchange interactions induced 
by the farther-neighbor hoppings play minor roles in
stabilizing the AFB order.

Another possible origin of the farther-neighbor magnetic interactions
is Ruderman-Kittel-Kasuya-Yosida (RKKY) interactions, which are mainly
induced by {the} Hund's rule {coupling}.
To examine the roles of {the} Hund's rule {coupling} as well as {of} the RKKY interactions,
we artificially change the magnitude of {the} Hund's rule coupling by introducing 
the scaling parameter
$\lambda_{J}$, where $\lambda_{J}$ scales
on-site exchange interactions as $J \rightarrow \lambda_{J} J$.
Figure~\ref{FeTe_energy_Jtcut}(b) shows the ground{-}state energy per site of several AF orders
in FeTe as a function of $\lambda_{J}$.
The AFB becomes more stable as $J$ becomes larger.
This result indicates that the RKKY-type interactions 
play key roles in stabilizing the AFB order.
Since scattering decreases {with} the magnetic ordering,
the resistivity decreases well below the magnetic transition temperature~\cite{MizuguchiJPSJ10}
{particularly} along the ($\pi$/2,$-\pi$/2) ferromagnetic line~\cite{Jiang13,LiuPRB15}.
The structural  transition~\cite{Martinelli10} would enhance the energy stabilization
through the changes in transfer parameters.
{The} couplings between 
itinerant electrons and localized spins {resulting in the} RKKY interactions
{were examined on phenomenological levels}~\cite{Yin_PRL,Akbari13,Ducatman14}.

\begin{figure}[tb]
\centering 
\includegraphics[clip,width=0.4\textwidth ]{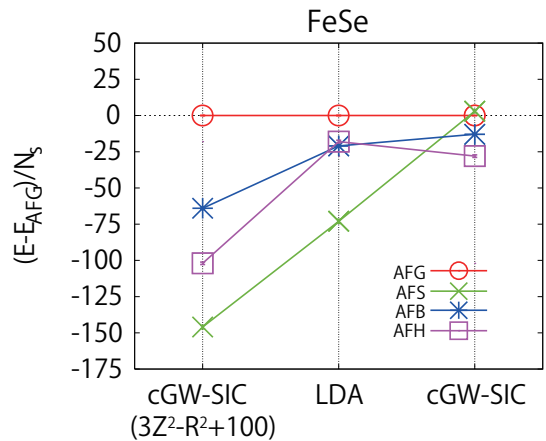} 
\caption{
{
(Color online) Ground{-}state energy per site of AF in the mVMC for FeSe with $t^{\text{LDA}}$ and with $t^{\text{cGW-SIC}}$ (in meV).
The left{-}hand data is {obtained by using} the cGW-SIC {calculation, but} the on-site potential of the $3Z^2-R^2$ orbital is increased by 100 meV {artificially}.
{The} system size is $N_{\text{s}}=8\times 8$.
}
}
\label{AF_FeSe_energy}
\end{figure}

Now, we discuss the magnetic properties {of} FeSe.
As in the case of FeTe,
the AFS phase becomes the ground state in the LDA calculation for FeSe, 
as shown in Fig.~\ref{AF_FeSe_energy}.
For FeSe, we only show the results of 
the largest system size ($N_{s}=8\times8$),
because the {size dependence} of the 
energy is not smooth in FeSe,
while we estimate the thermodynamic limit to be not far from the result for $N_{s}=8\times8$.
By eliminating the double counting of the self-energy,
we find that
the AFH phase has the lowest energy, which is consistent with previous {\it ab initio} calculations~\cite{Cao15,Glasbrenner}.
What is more characteristic of FeSe is that
the AFH and AFB phases win the AFS state and 
they have nearly the same energy, i.e.,
the difference in the energy between the AFH and AFB is {on} the order of $10$ meV {and the energies of the AFS and AFG phases are also close within 25 meV}.
{Such degeneracies were not properly captured in the previous DFT-GGA studies~\cite{Cao15,Glasbrenner}, where the energies of the AFB and {AFG} phases are much higher.}
Because of the limitation of the system sizes and the scattered size dependence of energy 
for FeSe, 
the energy stability of FeSe has a 
large uncertainty with {an} energy scale of {$10$ meV}.
Thus, this degeneracy does not contradict the experimental results 
that the magnetic order is not observed {despite} its large local moment,
since it is probable that the short-range order 
of the AFB and AFH (and AFS as well) coexist 
as {a} mixed or fluctuating phase, which does not allow the development of the 
long-range order above the superconducting transition temperature.

It is naively expected that the superposition of the 
degenerate magnetic orders {will induce} spin-glass-like states.
However, the {spin-glass-like} states are not consistent with
{the} existence of small but clear Fermi surfaces~\cite{TerashimaPRB2014,ShimojimaPRB2014} 
In the magnetically degenerate region,
it is plausible that an unconventional superconducting phase becomes stable.
By examining the stability of the superconducting phase by extending {the} present analysis for the normal states,
the origin of small Fermi surfaces and {the} resultant BEC-like superconductivity~\cite{Kasahara2014} will be clarified.
Further theoretical {investigation of} such a possibility is left for future study.
It is also {interesting} to determine the shapes of the Fermi surfaces
from jumps in momentum distributions by performing the calculations for larger system sizes.

The energy of the AFG phase is relatively low in FeSe compared with {that} {in} FeTe. 
The reason why the energy of the AFG phase {is} low is that 
the occupation number of the $XY$ orbital having {a} very weak frustration {is} also close to half filling ($1.10$)
in addition to those of the $YZ/ZX$ and $X^2-Y^2$ having stronger frustrations.
Note that the AFG order is more {stable} if the unfrustrated orbitals {are} closer to half filling.
This reduction of the $XY$ filling originates from the remarkable decrease in the on-site potential of the $3Z^2-R^2$ orbital in FeSe{;
thus,} the filling of the other orbitals must decrease.
However, the $YZ/ZX$ and $X^2-Y^2$ orbitals are already pinned near half filling.
This drives the filling of the $XY$ orbital closer to half filling.
Because the $XY$ orbital partially resolves the frustration in comparison to FeTe and the energy of the AFG phase {becomes} lower,
FeSe shows {an} accidental degeneracy of several phases.
More precisely, when 
{the three orbitals ($YZ$, $ZX$, and $X^2 - Y^2$) could become close to half filling}
by an appropriate on-site potential,
the AFS, AFB, and AFH phases
may have similar energies.
If the $XY$ orbital in addition {could become} close to half filling, the AFG phase joins in this degeneracy.

As we will show later, the degeneracy may be lifted easily under pressure.
The difference {in} the crystal field splitting from FeTe yields the unique energetic degeneracy of the AF orders {in FeSe}.

Lastly, we examine the robustness of the degeneracy in FeSe.
By comparing the results of the LDA and cGW-SIC models,
it is speculated that 
the stability of the AF phases 
largely depends on the on-site potential
of the $3Z^2-R^2$ orbital in iron chalcogenides.
To {determine} the roles of the on-site potential directly,
we {monitor the effect of} the on-site potential of the $3Z^2-R^2$ in the cGW-SIC model for FeSe.
As a result, we 
find that the degeneracy is easily lifted and 
{that} the AFS becomes the ground state as shown in 
Fig.~\ref{AF_FeSe_energy}.
This result can be understood by considering the fact that the occupation number of the $YZ/ZX$ orbitals
{deviates} from the half filling{,} and {that} {only} the $X^2-Y^2$ orbital,
which stabilizes the AFS states through large next-nearest {transfer},
{stays near half filling and} becomes {the sole {driving force}} of the magnetism.
This mechanism is hardly captured by weak-coupling theories~\cite{Suzuki11}.
{Although} not so easy, experimental
control of the one-body potential of the $3Z^2-R^2$ orbital
will reveal the origin of the degeneracy in FeSe.
The magnetic transition under pressure~\cite{Bendele10,TerashimaJPSJ15} would be related to the degeneracy of the AF phases.
In the {present} {\it ab initio} study using the cGW-SIC model of FeSe under 4 GPa,  such a degeneracy is lifted and it predicts that the AFH phase becomes stable{, while the AFS phase is competing with it}~\cite{SM2}.

In this Letter, we have shown that 
the elimination of {the} double counting in the exchange-correlation 
energy is essential {for reproducing} the AFB order observed in FeTe.
The stability of the AFB in FeTe is insensitive to farther-neighbor 
{transfer} but sensitive to the strength of {the} Hund's rule coupling.
{These} results indicate that the nesting in Fermi surfaces and superexchange interactions 
play minor {roles} but {that} {the} RKKY interactions induced by {the} Hund's rule coupling 
play a key role in stabilizing the AFB order.
We have also found that several magnetic orders are energetically
degenerate in FeSe.
This degeneracy is the possible origin of the absence of
magnetic order in FeSe. 
It is shown that this degeneracy can be lifted 
by changing the chemical potential of the $3Z^2-R^2$ orbital.
Our analysis provides a firm theoretical basis for understanding 
the characteristic magnetism in iron chalcogenides 
and offers a clue {to} understanding other
{unique} behaviors observed in {the} iron chalcogenides
such as the possible high-temperature superconductivity in
a single layer of FeSe~\cite{Qing_2012}.


We would like to thank Kazuma Nakamura for discussions on the dimensional downfolding method
{and Taichi Terashima for discussions on FeSe under pressure.}
This work has been supported by Grants-in-Aid for Scientific Research from the Ministry of Education, Culture, Sports,
Science and Technology of Japan (MEXT) under grant numbers 22104010 and 22340090.
This work has also been financially supported by MEXT HPCI Strategic Programs for Innovative Research (SPIRE) and Computational
Materials Science Initiative (CMSI).
We also acknowledge K computer at RIKEN Advanced Institute for Computational
Science (AICS) under grant {numbers} hp120043, hp120283, hp130007{,} hp140215{, and hp150211}.


\bibliographystyle{jpsj_mod}

\clearpage
\noindent
{\Large Supplemental Materials}

\renewcommand{\theequation}{S.\arabic{equation}}
\setcounter{equation}{0}
\renewcommand{\tablename}{Table S}
\setcounter{table}{0}
\renewcommand{\figurename}{Fig. S}
\setcounter{figure}{0}

\noindent
\section*{ S.1 Values of Parameters of Low-Energy Effective Model}

\begin{figure}[tb]
\centering  
\includegraphics[width=0.45\textwidth ]{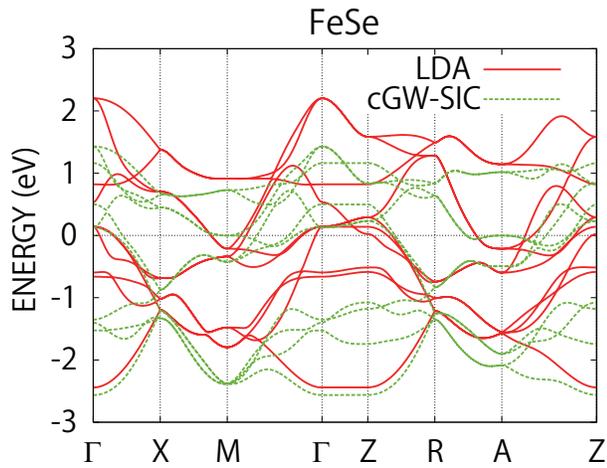}  
\caption{
(Color online) Electronic band structures of the Fe $3d$ {MLWFs} of FeSe in the LDA [(red) solid line]
 and the cGW-SIC [(green) dashed line].
The Fermi energy is set to zero.
}
\label{FeSe_band}
\end{figure}

\begin{figure}[ptb]
\begin{center} 
\includegraphics[width=0.45\textwidth ]{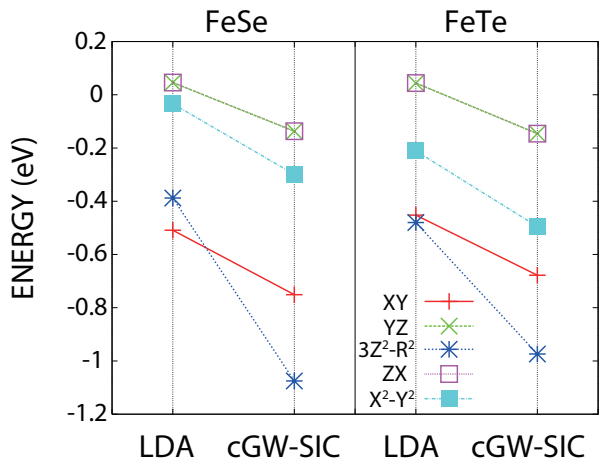} 
\end{center} 
\caption{(Color online) On-site potential of Wannier orbitals for FeSe and FeTe {obtained by} the LDA and cGW-SIC {calculations}.
}
\label{ton}
\end{figure}

In this supplemental material, we show the values of the parameters of the {\it ab initio} low-energy effective model. 
Table S~\ref{W_FeSe} (\ref{W_FeTe}) is the three-dimensional effective interaction of the maximally localized Wannier functions ({MLWFs}) originating from the Fe $3d$ orbitals of FeSe (FeTe)~\cite{S_hirayama13}.
We obtain the two-dimensional effective interaction by uniformly subtracting 0.6 (0.4) eV from the three-dimensional effective interaction of FeSe (FeTe)
following Refs.~\citen{S_nakamura10} and \citen{S_nakamurap}.
Table S~\ref{t_LDA} is the transfer integral of the {MLWFs} of FeSe and FeTe {obtained by} the local-density-approximation (LDA) {calculation}.
We also show the transfer integral of the {MLWFs} in the constrained GW with the self-interaction correction (cGW-SIC) model in Table S~\ref{t_cGW-SIC}.
Figure S~\ref{FeSe_band} is the corresponding band structure of FeSe.
The on-site potential of each model is shown in  Fig. S~\ref{ton}.
The detail of the cGW method is explained in Refs.~\citen{S_hirayama13} and \citen{S_aryasetiawan09}.

\begin{figure}[tb]
\centering 
\includegraphics[clip,width=0.4\textwidth ]{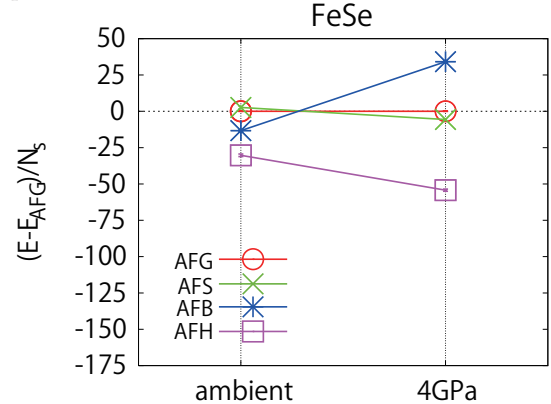} 
\caption{
(Color online) Ground state energy per site of AF in the mVMC for FeSe at ambient pressure and at 4.0 GPa with $t^{\text{cGW-SIC}}$ (in meV).
System size is $N_{\text{s}}=8\times 8$.
}
\label{AF_FeSe_energy_4GPa}
\end{figure} 

\section*{ S.2 Magnetism of FeSe under Pressure}

We calculate the magnetism of FeSe under pressure as well as that at ambient pressure. 
Figure S~\ref{AF_FeSe_energy_4GPa} shows the energy per site {by} the cGW-SIC {calculation for} FeSe at 4.0 GPa in the experimental geometry~\cite{S_margadonna09}.
{Although the AFS-type short-ranged fluctuations were observed~\cite{S_rahn15,S_wangX} at ambient pressure, the} unique energetic degeneracy at ambient pressure is lifted under pressure, and we predict that the AFH phase becomes {slightly more} stable {than the AFS}.
The experimental verification would be desired.

\begin{table*}[htb] 
\caption{
Bare and effective three-dimensional Coulomb interactions between two electrons for all the combinations of Fe $3d$ orbitals in FeSe (in eV).
Here, $v$ and $J_{v}$ represent the bare on-site and exchange Coulomb interactions, respectively. 
The static limit
of the effective on-site and exchange Coulomb interactions are denoted by $U$ and $J$, respectively. 
}
\ 
\label{W_FeSe} 
\scalebox{0.85}[0.85]{
\begin{tabular}{c|ccccc|ccccc} 
\hline \hline \\ [-8pt]
FeSe           &      &      &     $v$     &      &               &      &      &      $U$    &      &          \\ [+1pt]
\hline \\ [-8pt]
               & $XY$ & $YZ$ & $3Z^{2}-R^{2}$ & $ZX$ & $X^{2}-Y^{2}$ & $XY$ & $YZ$ & $3Z^{2}-R^{2}$ & $ZX$ & $X^{2}-Y^{2}$ \\ 
\hline \\ [-8pt] 
$XY$           & 18.65 & 16.50 & 17.28 & 16.50 & 16.51  & 4.51 & 3.19 & 3.21 & 3.19 & 3.48 \\
$YZ$           & 16.50 & 16.97 & 17.05 & 15.70 & 15.30  & 3.19 & 4.11 & 3.53 & 3.02 & 2.98 \\
$3Z^{2}-R^{2}$ & 17.28 & 17.05 & 19.09 & 17.05 & 16.00  & 3.21 & 3.53 & 4.68 & 3.53 & 3.01 \\
$ZX$           & 16.50 & 15.70 & 17.05 & 16.97 & 15.30  & 3.19 & 3.02 & 3.53 & 4.11 & 2.98 \\
$X^{2}-Y^{2}$  & 16.51 & 15.30 & 16.00 & 15.30 & 15.88  & 3.48 & 2.98 & 3.01 & 2.98 & 3.77 \\
\hline \hline \\ [-8pt]
               &      &      &     $J_{v}$     &      &               &      &      &      $J$    &      &          \\ [+1pt]
\hline \\ [-8pt]
               & $XY$ & $YZ$ & $3Z^{2}-R^{2}$ & $ZX$ & $X^{2}-Y^{2}$ & $XY$ & $YZ$ & $3Z^{2}-R^{2}$ & $ZX$ & $X^{2}-Y^{2}$ \\ 
\hline \\ [-8pt] 
$XY$           &      & 0.66 & 0.79 & 0.66 & 0.34 &      & 0.57 & 0.69 & 0.57 & 0.32  \\
$YZ$           & 0.66 &      & 0.46 & 0.56 & 0.61 & 0.57 &      & 0.42 & 0.48 & 0.53  \\
$3Z^{2}-R^{2}$ & 0.79 & 0.46 &      & 0.46 & 0.75 & 0.69 & 0.42 &      & 0.42 & 0.62  \\
$ZX$           & 0.58 & 0.56 & 0.46 &      & 0.61 & 0.57 & 0.48 & 0.42 &      & 0.53  \\
$X^{2}-Y^{2}$  & 0.34 & 0.61 & 0.75 & 0.61 &      & 0.32 & 0.53 & 0.62 & 0.53 &       \\
\hline
\hline 
\end{tabular}
}
\end{table*}

\begin{table*}[htb] 
\caption{
Bare and effective three-dimensional Coulomb interactions between two electrons for all the combinations of Fe $3d$ orbitals in FeTe (in eV).
Here, $v$ and $J_{v}$ represent the bare on-site and exchange Coulomb interactions, respectively. 
The static limit
of the effective on-site and exchange Coulomb interactions are denoted by $U$ and $J$, respectively. 
}
\ 
\label{W_FeTe} 
\scalebox{0.85}[0.85]{
\begin{tabular}{c|ccccc|ccccc} 
\hline \hline \\ [-8pt]
FeTe           &      &      &     $v$     &      &               &      &      &      $U$    &      &          \\ [+1pt]
\hline \\ [-8pt]
               & $XY$ & $YZ$ & $3Z^{2}-R^{2}$ & $ZX$ & $X^{2}-Y^{2}$ & $XY$ & $YZ$ & $3Z^{2}-R^{2}$ & $ZX$ & $X^{2}-Y^{2}$ \\ 
\hline \\ [-8pt] 
$XY$           & 17.08 & 15.05 & 16.20 & 15.05 & 16.26  & 3.46 & 2.30 & 2.36 & 2.30 & 2.80 \\
$YZ$           & 15.05 & 15.36 & 15.87 & 14.24 & 14.94  & 2.30 & 3.08 & 2.62 & 2.15 & 2.30 \\
$3Z^{2}-R^{2}$ & 16.20 & 15.87 & 18.25 & 15.87 & 16.08  & 2.36 & 2.62 & 3.73 & 2.62 & 2.35 \\
$ZX$           & 15.05 & 14.24 & 15.87 & 15.36 & 14.94  & 2.30 & 2.15 & 2.62 & 3.08 & 2.30 \\
$X^{2}-Y^{2}$  & 16.32 & 14.94 & 16.08 & 14.94 & 16.77  & 2.82 & 2.30 & 2.36 & 2.30 & 3.39\\
\hline \hline \\ [-8pt]
               &      &      &     $J_{v}$     &      &               &      &      &      $J$    &      &          \\ [+1pt]
\hline \\ [-8pt]
               & $XY$ & $YZ$ & $3Z^{2}-R^{2}$ & $ZX$ & $X^{2}-Y^{2}$ & $XY$ & $YZ$ & $3Z^{2}-R^{2}$ & $ZX$ & $X^{2}-Y^{2}$ \\ 
\hline \\ [-8pt] 
$XY$           &      & 0.59 & 0.73 & 0.59 & 0.33 &      & 0.49 & 0.62 & 0.49 & 0.31 \\
$YZ$           & 0.59 &      & 0.42 & 0.49 & 0.59 & 0.49 &      & 0.37 & 0.40 & 0.49 \\
$3Z^{2}-R^{2}$ & 0.73 & 0.42 &      & 0.42 & 0.74 & 0.62 & 0.37 &      & 0.37 & 0.62 \\
$ZX$           & 0.59 & 0.49 & 0.42 &      & 0.59 & 0.49 & 0.40 & 0.37 &      & 0.49 \\
$X^{2}-Y^{2}$  & 0.33 & 0.59 & 0.74 & 0.59 &      & 0.31 & 0.49 & 0.62 & 0.49 &      \\
\hline
\hline 
\end{tabular}
}
\end{table*}

\begin{table*}[ptb] 
\caption{
Transfer integral for the $3d$ orbitals of the Fe sites in the FeSe and FeTe, $t^{\text{LDA}}_{mn}(R_x, R_y, R_z)$, 
where $t^{\text{LDA}}$ is the expectation value of the Kohn-Sham Hamiltonian for the Wannier function : $t^{\text{LDA}}=\langle \phi | \mathcal{H}^{\text{LDA}} |\phi \rangle$, $m$ and $n$ denote symmetries of the $3d$ orbitals, and the axis of $(R_x, R_y, R_z)$ is taken along the Fe-Se/Te directions.
Units are given in meV.  
} 
\
\scalebox{0.75}[0.75]{ 
\begin{tabular}{c|rrrrrrr|rrr} 
\hline \hline \\ [-4pt]
  FeSe    \\ [+2pt] 
\hline \\ [-4pt]
$(m, n)$ $\backslash$ $\bm{R}$ 
& \big[$0,0,0$\big] 
& \big[$\frac{1}{2},-\frac{1}{2},0$\big] 
& \big[$1,0,0$\big] 
& \big[$1,-1,0$\big] 
& \big[$\frac{3}{2},-\frac{1}{2},0$\big]
& \big[$0,0,\frac{c}{a}$\big] 
& \big[$\frac{1}{2},-\frac{1}{2},\frac{c}{a}$\big]
& $\sigma_{Y}$
& $I$
& $\sigma^{L}$ \\ [+4pt]
\hline \\ [-8pt]             
$(XY,XY)$                    & -509 &   -410 &    -70 &    -11 &      3 &    -25 &      6 & $+$ & $+$ &      $+$ \\ 
$(XY,YZ)$                    &    0 &    273 &    131 &     -9 &     -6 &      0 &     -9 & $+$ & $-$ & $-$(1,4) \\
$(XY,3Z^{2}-R^{2})$          &    0 &   -347 &      0 &     22 &     -8 &      0 &     11 & $-$ & $+$ &      $+$ \\ 
$(XY,ZX)$                    &    0 &    273 &      0 &     -9 &     18 &      0 &     -3 & $-$ & $-$ & $-$(1,2) \\
$(XY,X^{2}-Y^{2})$           &    0 &      0 &      0 &      0 &     -9 &      0 &     -4 & $-$ & $+$ &      $-$ \\ 
$(YZ,YZ)$                    &   46 &    197 &    128 &    -17 &     -8 &      8 &     27 & $+$ & $+$ &    (4,4) \\ 
$(YZ,3Z^{2}-R^{2})$          &    0 &   -119 &      0 &      7 &      2 &      0 &     11 & $-$ & $-$ & $-$(4,3) \\ 
$(YZ,ZX)$                    &    0 &    127 &      0 &    -23 &    -19 &      0 &     12 & $-$ & $+$ &    (4,2) \\ 
$(YZ,X^{2}-Y^{2})$           &    0 &    223 &      0 &      1 &     -3 &      0 &     20 & $-$ & $-$ &    (4,5) \\ 
$(3Z^{2}-R^{2},3Z^{2}-R^{2})$& -388 &     -4 &    -15 &    -14 &     -6 &    -23 &     -9 & $+$ & $+$ &      $+$ \\ 
$(3Z^{2}-R^{2},ZX)$          &    0 &    119 &    199 &     -7 &    -13 &      0 &    -10 & $+$ & $-$ & $-$(3,2) \\ 
$(3Z^{2}-R^{2},X^{2}-Y^{2})$ &    0 &      0 &   -115 &      0 &      1 &     -8 &     -6 & $+$ & $+$ &      $-$ \\ 
$(ZX,ZX)$                    &   46 &    197 &    335 &    -17 &     13 &      8 &      0 & $+$ & $+$ &    (2,2) \\ 
$(ZX,X^{2}-Y^{2})$           &    0 &   -223 &     82 &     -1 &    -15 &      0 &      7 & $+$ & $-$ &    (2,5) \\ 
$(X^{2}-Y^{2},X^{2}-Y^{2})$  &  -34 &    -56 &     93 &      0 &     17 &    -28 &      4 & $+$ & $+$ &      $+$ \\ 
\hline \hline 
  FeTe    \\ [+2pt] 
\hline \\ [-4pt]
$(m, n)$ $\backslash$ $\bm{R}$ 
& \big[$0,0,0$\big] 
& \big[$\frac{1}{2},-\frac{1}{2},0$\big] 
& \big[$1,0,0$\big] 
& \big[$1,-1,0$\big] 
& \big[$\frac{3}{2},-\frac{1}{2},0$\big]
& \big[$0,0,\frac{c}{a}$\big] 
& \big[$\frac{1}{2},-\frac{1}{2},\frac{c}{a}$\big]
& $\sigma_{Y}$
& $I$
& $\sigma^{L}$ \\ [+4pt] 
\hline \\ [-8pt]             
$(XY,XY)$                    &  -452 &   -378 &    -11 &    -41 &     -1 &    -31 &     12 & $+$ & $+$ &      $+$ \\ 
$(XY,YZ)$                    &       0 &    237 &    109 &      3 &     -6 &      0 &    -12 & $+$ & $-$ & $-$(1,4) \\
$(XY,3Z^{2}-R^{2})$          &      0 &   -336 &      0 &     33 &     -9 &      0 &     21 & $-$ & $+$ &      $+$ \\ 
$(XY,ZX)$                    &     0 &    237 &      0 &      3 &     35 &      0 &     -1 & $-$ & $-$ & $-$(1,2) \\
$(XY,X^{2}-Y^{2})$           &       0 &      0 &      0 &      0 &    -14 &      0 &      3 & $-$ & $+$ &      $-$ \\ 
$(YZ,YZ)$                    &   44 &    156 &    103 &    -15 &    -12 &     13 &     37  & $+$ & $+$ &    (4,4) \\ 
$(YZ,3Z^{2}-R^{2})$          &     0 &   -122 &      0 &     17 &      6 &      0 &     11 & $-$ & $-$ & $-$(4,3) \\ 
$(YZ,ZX)$                    &     0 &    101 &      0 &    -27 &    -25 &      0 &     14  & $-$ & $+$ &    (4,2) \\ 
$(YZ,X^{2}-Y^{2})$           &      0 &    178 &      0 &      0 &    -10 &      0 &     22 & $-$ & $-$ &    (4,5) \\ 
$(3Z^{2}-R^{2},3Z^{2}-R^{2})$& -480 &    -73 &    -53 &      3 &      8 &    -67 &    -23 & $+$ & $+$ &      $+$ \\ 
$(3Z^{2}-R^{2},ZX)$          &    0 &    122 &    198 &    -17 &    -17 &      0 &    -32 & $+$ & $-$ & $-$(3,2) \\ 
$(3Z^{2}-R^{2},X^{2}-Y^{2})$ &      0 &      0 &    -29 &      0 &     -6 &     30 &    -30  & $+$ & $+$ &      $-$ \\ 
$(ZX,ZX)$                    &    44 &    156 &    300 &    -15 &     42 &     13 &      9  & $+$ & $+$ &    (2,2) \\ 
$(ZX,X^{2}-Y^{2})$           &     0 &   -178 &    136 &      0 &    -23 &      0 &     26  & $+$ & $-$ &    (2,5) \\ 
$(X^{2}-Y^{2},X^{2}-Y^{2})$  & -211 &     66 &     52 &      9 &     12 &     16 &    -24 & $+$ & $+$ &      $+$ \\ 
\hline \hline 
\end{tabular}
}
\label{t_LDA} 
\end{table*}

\begin{table*}[ptb] 
\caption{Transfer integral for the $3d$ orbitals of the Fe sites in the FeSe and FeTe, $t^{\text{cGW-SIC}}_{mn}(R_x, R_y, R_z)$, 
where $t^{\text{cGW-SIC}}$ is the transfer integral without double-counting :
$t^{\text{cGW-SIC}}=\langle \phi | \mathcal{H}^{\text{cGW-SIC}} |\phi \rangle$,
$m$ and $n$ denote symmetries of the $3d$ orbitals, and the axis of $(R_x, R_y, R_z)$ is taken along the Fe-Se/Te directions.
Units are given in meV.
} 
\
\scalebox{0.75}[0.75]{ 
\begin{tabular}{c|rrrrrrr|rrr} 
\hline \hline \\ [-4pt]
  FeSe    \\ [+2pt] 
\hline \\ [-4pt]
$(m, n)$ $\backslash$ $\bm{R}$ 
& \big[$0,0,0$\big] 
& \big[$\frac{1}{2},-\frac{1}{2},0$\big] 
& \big[$1,0,0$\big] 
& \big[$1,-1,0$\big] 
& \big[$\frac{3}{2},-\frac{1}{2},0$\big]
& \big[$0,0,\frac{c}{a}$\big] 
& \big[$\frac{1}{2},-\frac{1}{2},\frac{c}{a}$\big]
& $\sigma_{Y}$
& $I$
& $\sigma^{L}$ \\ [+4pt]
\hline \\ [-8pt]             
$(XY,XY)$                    & -751 &   -466 &    -20 &     24 &      3 &    -38 &      6 & $+$ & $+$ &      $+$ \\ 
$(XY,YZ)$                    &    0 &    201 &    106 &    -26 &     -3 &      0 &     -9 & $+$ & $-$ & $-$(1,4) \\
$(XY,3Z^{2}-R^{2})$          &     0 &   -391 &      0 &     27 &     -3 &      0 &      8 & $-$ & $+$ &      $+$ \\ 
$(XY,ZX)$                    &   0 &    201 &      0 &    -26 &     -9 &      0 &     -4 & $-$ & $-$ & $-$(1,2) \\
$(XY,X^{2}-Y^{2})$           &    0 &      0 &      0 &      0 &      6 &      0 &    -10 & $-$ & $+$ &      $-$ \\ 
$(YZ,YZ)$                    &  -137 &     97 &    148 &    -35 &    -15 &      7 &     27 & $+$ & $+$ &    (4,4) \\ 
$(YZ,3Z^{2}-R^{2})$          &    0 &    -94 &      0 &     22 &     -1 &      0 &      8 & $-$ & $-$ & $-$(4,3) \\ 
$(YZ,ZX)$                    &    0 &    199 &      0 &    -37 &    -36 &      0 &     16 & $-$ & $+$ &    (4,2) \\ 
$(YZ,X^{2}-Y^{2})$           &     0 &    204 &      0 &     -1 &      7 &      0 &     17 & $-$ & $-$ &    (4,5) \\ 
$(3Z^{2}-R^{2},3Z^{2}-R^{2})$&  -1075 &    -67 &    -76 &      5 &     11 &     -4 &      3 & $+$ & $+$ &      $+$ \\ 
$(3Z^{2}-R^{2},ZX)$          &     0 &     94 &    195 &    -22 &    -16 &      0 &     -3  & $+$ & $-$ & $-$(3,2) \\ 
$(3Z^{2}-R^{2},X^{2}-Y^{2})$ &     0 &      0 &    -35 &      0 &    -13 &    -22 &      7 & $+$ & $+$ &      $-$ \\ 
$(ZX,ZX)$                    &    -137 &     97 &    263 &    -35 &     25 &      7 &     -2 & $+$ & $+$ &    (2,2) \\ 
$(ZX,X^{2}-Y^{2})$           &     0 &   -204 &    147 &      1 &    -39 &      0 &      0  & $+$ & $-$ &    (2,5) \\ 
$(X^{2}-Y^{2},X^{2}-Y^{2})$  &   -299 &    186 &     19 &    -20 &      8 &    -35 &     17 & $+$ & $+$ &      $+$ \\ 
\hline \hline 
  FeTe    \\ [+2pt] 
\hline \\ [-4pt]
$(m, n)$ $\backslash$ $\bm{R}$ 
& \big[$0,0,0$\big] 
& \big[$\frac{1}{2},-\frac{1}{2},0$\big] 
& \big[$1,0,0$\big] 
& \big[$1,-1,0$\big] 
& \big[$\frac{3}{2},-\frac{1}{2},0$\big]
& \big[$0,0,\frac{c}{a}$\big] 
& \big[$\frac{1}{2},-\frac{1}{2},\frac{c}{a}$\big]
& $\sigma_{Y}$
& $I$
& $\sigma^{L}$ \\ [+4pt] 
\hline \\ [-8pt]             
$(XY,XY)$                    &  -678 &   -410 &     52 &      5 &     -2 &     23 &    -13 & $+$ & $+$ &      $+$ \\ 
$(XY,YZ)$                    &     0 &    133 &     79 &    -11 &     -1 &      0 &     11 & $+$ & $-$ & $-$(1,4) \\
$(XY,3Z^{2}-R^{2})$          &     0 &   -340 &      0 &     26 &     -8 &      0 &     11 & $-$ & $+$ &      $+$ \\ 
$(XY,ZX)$                    &    0 &    133 &      0 &    -11 &     -2 &    -26 &     10  & $-$ & $-$ & $-$(1,2) \\
$(XY,X^{2}-Y^{2})$           &     0 &      0 &      0 &      0 &     -1 &      0 &     16 & $-$ & $+$ &      $-$ \\ 
$(YZ,YZ)$                    &   -146 &     42 &    120 &    -28 &    -17 &     11 &     15  & $+$ & $+$ &    (4,4) \\ 
$(YZ,3Z^{2}-R^{2})$          &     0 &    -88 &      0 &     24 &      6 &    -23 &     14 & $-$ & $-$ & $-$(4,3) \\ 
$(YZ,ZX)$                    &     0 &    186 &      0 &    -35 &    -46 &      0 &     16  & $-$ & $+$ &    (4,2) \\ 
$(YZ,X^{2}-Y^{2})$           &    0 &    139 &      0 &      1 &     -2 &     12 &     14 & $-$ & $-$ &    (4,5) \\ 
$(3Z^{2}-R^{2},3Z^{2}-R^{2})$& -974 &   -114 &   -114 &     21 &     26 &     -1 &     -1  & $+$ & $+$ &      $+$ \\ 
$(3Z^{2}-R^{2},ZX)$          &    0 &     88 &    181 &    -24 &    -12 &      0 &     -4 & $+$ & $-$ & $-$(3,2) \\ 
$(3Z^{2}-R^{2},X^{2}-Y^{2})$ &    0 &      0 &     49 &      0 &    -24 &     -1 &      2 & $+$ & $+$ &      $-$ \\ 
$(ZX,ZX)$                    &  -146 &     42 &    213 &    -28 &     51 &     22 &     13 & $+$ & $+$ &    (2,2) \\ 
$(ZX,X^{2}-Y^{2})$           &    0 &   -139 &    147 &     -1 &    -30 &      0 &      8 & $+$ & $-$ &    (2,5) \\ 
$(X^{2}-Y^{2},X^{2}-Y^{2})$  &   -495 &    292 &    -47 &     -9 &     20 &      5 &    -13 & $+$ & $+$ &      $+$ \\ 
\hline \hline 
\end{tabular}
}
\label{t_cGW-SIC} 
\end{table*}


\end{document}